# Current-induced motion of twisted skyrmions

Chendong Jin[1], Chunlei Zhang[1], Chengkun Song[1], Jinshuai Wang[1], Haiyan Xia[1], Yunxu Ma[1], Jianing Wang[1], Yurui Wei[1], Jianbo Wang[1,2] and Qingfang Liu[1,*]

[1]*Key Laboratory for Magnetism and Magnetic Materials of the Ministry of Education, Lanzhou University, Lanzhou 730000, People's Republic of China.*

[2]*Key Laboratory for Special Function Materials and Structural Design of the Ministry of the Education, Lanzhou University, Lanzhou 730000, People's Republic of China.*

**Abstract**

Twisted skyrmions, whose helicity angles are different from that of Bloch skyrmions and Néel skyrmions, have already been demonstrated in experiments recently. In this work, we first contrast the magnetic structure and origin of the twisted skyrmion with other three types of skyrmion including Bloch skyrmion, Néel skyrmion and antiskyrmion. Following, we investigate the dynamics of twisted skyrmions driven by the spin transfer toque (STT) and the spin Hall effect (SHE) by using micromagnetic simulations. It is found that the spin Hall angle of the twisted skyrmion is related to the dissipative force tensor and the Gilbert damping both for the motions induced by the STT and the SHE, especially for the SHE induced motion, the skyrmion Hall angle depends substantially on the skyrmion helicity. At last, we demonstrate that the trajectory of the twisted skyrmion can be controlled in a two dimensional plane with a Gilbert damping gradient. Our results provide the understanding of current-induced motion of twisted skyrmions, which may contribute to the applications of skyrmion-based racetrack memories.

___________________________________

*Corresponding author: Qingfang Liu, liuqf@lzu.edu.cn

**Introduction**

It has been recognized that the spin-polarized current-induced the motion and reversal of magnetic structures arises as a result of the spin transfer torque (STT) effect[1-4], which has attracted large interests due to the fundamental physics and potential applications in spintronic devices, such as magnetic random access memories (MRAMs)[5, 6], racetrack memories[7, 8], nano-oscillators[9-11] and logic devices[12-14]. Recently, it has been reported that the spin Hall effect (SHE)[15, 16], generated by the pure spin currents flowed from the heavy metal substrate due to the strong spin-orbit coupling at the interface of ferromagnet/heavy-metal, is an alternative efficient method to manipulate the magnetization dynamics in magnetic materials[17-20]. Compared with the STT, the SHE does not require currents flow through the magnetic layer, and then reducing the Joule heat and electromigration, i.e., avoiding the restricted effect of large current density in traditional STT devices[21].

Magnetic skyrmions are chiral spin magnetization structures with topological properties and can be divided into the following types according to different types of Dzyaloshinskii-Moriya interaction (DMI)[22-26]: (i) Bloch skyrmmions are first discovered in bulk non-centrosymmetric B20-type lattice structures such as MnSi[27], FeCoSi[28-30], and FeGe[31, 32] due to the presence of bulk DMI; (ii) Néel skyrmions are observed in multilayered ultrathin films lacking inversion symmetry with strong spin-orbit coupling like Ir(111)/Fe[33], Ta/CoFeB[34] and Pt/Co[35] due to the presence of interfacial DMI; (iii) Antiskyrmions are reported in Heusler compounds such as MnPtSn[36] due to the presence of anisotropic DMI[37, 38]. Recently, at the interface of chiral bulk $Cu_2OSeO_3$ below a certain thickness, the so-called twisted skyrmions are demonstrated directly by the circularly polarized resonant elastic x-ray scattering, due to the breaking of translational symmetry at the surface of bulk ferromagnet[39, 40]. Up to now, the dynamics of twisted skyrmions driven by current have not been reported. Therefore, in this paper, on the basis of comparing the magnetic structure, origin and topological properties of the above four types of skyrmion, we focus on the dynamics of twisted skyrmions driven by the STT and the SHE and also analysis the simulation results by using Thiele's equations[41].

**Micromagnetic simulation details**

Our magnetic simulation results are performed by using the Object Oriented MicroMagnetic Framework (OOMMF) public code[42], which includes the additional modules for bulk DMI, interfacial DMI, anisotropic DMI and twisted DMI. The magnetization dynamics is described by numerically solving the Landau-Lifshitz-Gilbert (LLG) equation containing terms of the STT and the SHE[17, 20], as follow:

$$\frac{d\vec{m}}{dt} = -\gamma \vec{m} \times \vec{H}_{eff} + \alpha \vec{m} \times \frac{d\vec{m}}{dt} + \vec{\tau}_{STT} + \vec{\tau}_{SHE}, \tag{1}$$

where $\vec{m}$ is the unit vector of the local magnetization, $\gamma$ is the gyromagnetic ratio, $\alpha$ is the Gilbert damping, $\vec{H}_{eff}$ is the effective field including the exchange field, anisotropy field, demagnetization field and DMI effective field. The STT term is expressed as

$$\vec{\tau}_{STT} = v_s \vec{m} \times (\frac{\partial \vec{m}}{\partial x} \times \vec{m}) + \beta v_s (\vec{m} \times \frac{\partial \vec{m}}{\partial x}), \tag{2}$$

where $\beta$ is the non-adiabatic factor, and $v_s$ is the velocity of the conduction electrons with the form $v_s = \frac{\gamma \hbar P}{2\mu_0 e M_s} J$, where $J$ is the current density, $e$ is the electron charge, $P$ is the spin polarization, $\hbar$ is the reduced Planck constant, $\mu_0$ is the permeability of free space, and $M_s$ is the saturation magnetization. The electrons flowing toward $+x$ direction when $v_s > 0$. The SHE term is given by

$$\vec{\tau}_{SHE} = -\frac{\gamma \hbar}{2\mu_0 e M_s L} \theta_{SH} \vec{m} \times \vec{m} \times (\vec{z} \times \vec{j}_{HM}), \tag{3}$$

where $L$ is the thickness of the magnetic layer with the value of 1 nm, $\theta_{SH}$ is the spin-Hall angle of Pt substrate with the value of 0.07, $\vec{z}$ is the unit vectors of the surface normal direction, and $\vec{j}_{HM}$ is the current density injected into the heavy metal.

In order to eliminate the influence of the boundary effect on the size and dynamics of skyrmions, the 2D plane is assumed to $500 \times 500 \times 1$ nm$^3$ (length $\times$ width $\times$ thickness) with the mesh size of $1 \times 1 \times 1$ nm$^3$, and the initial position of the skyrmion is set in the center of the 2D plane. The material parameters are chosen similar to Ref. [8]: saturation magnetization $M_s = 580 \times 10^3$ A/m, exchange constant $A = 1.5 \times 10^{-11}$ J/m, perpendicular magnetic anisotropy constant

$K_u = 8 \times 10^5$ J/m$^3$, and DMI strength $D_{DMI} = 2.5 \sim 3.5 \times 10^{-3}$ J/m$^2$.

**Four types of skyrmions**

According to the different helicity of skyrmions, there are four types of skyrmions: Bloch skyrmion, Néel skyrmion, antiskyrmion and twisted skyrmion as shown in Figs. 1(a)–(d), respectively. Figures 1(e)–(h) display the corresponding spatial profiles of the local magnetization across the skyrmions. It can be seen that the $m_z$ of the four types of skyrmions are consistent, while the $m_x$ and $m_y$ of the four types of skyrmions are different, which again proves the different distribution of the in-plane magnetic moments of the four types of skyrmions. We emphasize that the distribution of the in-plane magnetic moments in the skyrmion structure is determined by the direction of the DMI vector, that is to say, the existence of the twisted skyrmion in this work is achieved by changing the DMI vector, which is much different as the reason that observed in the experiments. Figures 1(i)–(l) show the four types of DMIs: bulk DMI, interfacial DMI, anisotropic DMI and twisted DMI that promise the existence of the Bloch skyrmion, Néel skyrmion, antiskyrmion and twisted skyrmion, respectively. The four types of DMI considered in C4 symmetry can be written as:

$$E_{\text{BulkDMI}} = \frac{D}{2} \sum_i \vec{S}_i \times (\vec{S}_{i+\hat{x}} \times \hat{x} - \vec{S}_{i-\hat{x}} \times \hat{x} + \vec{S}_{i+\hat{y}} \times \hat{y} - \vec{S}_{i-\hat{y}} \times \hat{y}),$$

$$E_{\text{InterDMI}} = \frac{D}{2} \sum_i \vec{S}_i \times (-\vec{S}_{i+\hat{x}} \times \hat{y} + \vec{S}_{i-\hat{x}} \times \hat{y} + \vec{S}_{i+\hat{y}} \times \hat{x} - \vec{S}_{i-\hat{y}} \times \hat{x}),$$

$$E_{\text{AnisoDMI}} = \frac{D}{2} \sum_i \vec{S}_i \times (\vec{S}_{i+\hat{x}} \times \hat{y} - \vec{S}_{i-\hat{x}} \times \hat{y} + \vec{S}_{i+\hat{y}} \times \hat{x} - \vec{S}_{i-\hat{y}} \times \hat{x}),$$

$$E_{\text{TwisDMI}} = \frac{D}{2} \sum_i \vec{S}_i \times (\vec{S}_{i+\hat{x}} \times (\hat{x} - \hat{y}) + \vec{S}_{i-\hat{x}} \times (-\hat{x} + \hat{y}) + \vec{S}_{i+\hat{y}} \times (\hat{x} + \hat{y}) - \vec{S}_{i-\hat{y}} \times (\hat{x} + \hat{y}),$$

(4)

where $D$ is the DMI constant representing the DMI strength, $\vec{S}_i$ is the atomic moment unit vector, $\hat{x}$ and $\hat{y}$ are the unit vectors in the model.

**Topological properties of four types of skyrmions**

**A. Helicity, winding number and topological number**

In order to better understand the helicity and winding number of skyrmions, we use the two-dimensional polar coordinates to describe a general magnetic skyrmion structure, as shown in Fig. 2 which displays a Bloch skyrmion, as example, in the polar coordinates with azimuthal angle ($\varphi$) and radial coordinate ($\rho$). Therefore, the unit vector of the local magnetization $m_x$, $m_y$ and $m_z$ in the Cartesian coordinates can be written as[26, 43, 44]:

$$
\begin{aligned}
m_x &= \sin\theta(\rho)\cos\phi(\varphi), \\
m_y &= \sin\theta(\rho)\sin\phi(\varphi), \\
m_z &= \cos\theta(\rho),
\end{aligned}
\tag{5}
$$

where $\theta(\rho)$ is the radial profile of the perpendicular component of the magnetization, and from the center to the boundary, its value changes from $-0.5\pi$ to $0.5\pi$, $\phi(\varphi)$ is the angle between the magnetic moment and the radial coordinate. The vorticity of skyrmions is obtained by calculating the full turns of the transverse magnetic moments on the perimeter and is defined by the winding number[45] $W = \frac{1}{2\pi}\int_{\varphi=0}^{2\pi}\mathrm{d}\phi(\varphi)$. Therefore, the winding number $W = 1$ for twisted skyrmion, Bloch skyrmion and Néel skyrmion as shown in Fig. 3(a), and $W = -1$ for antiskyrmion as shown in Fig. 3(b). The helicity of a skyrmion is given by $\chi = \phi(\varphi) - W\cdot\varphi = \phi(\varphi = 0)$ with the value ranging form $-\pi$ to $\pi$, that is, for the Bloch skyrmion, $\chi = \pm 0.5\pi$; for Néel skyrmion, $\chi = 0$ or $\pi$; for twisted skyrmion, $\chi \neq \pm 0.5\pi$, 0 and $\pi$, and the helicity $\chi$ of the twisted skyrmion shown in Fig. 1(d) equals to $0.25\pi$; for antiskyrmion as shown in Fig. 1(c), $\chi = \pi$. The topological number $Q$ relates to the winding number and counts how many times the unit vector along the magnetic moment wraps the unit sphere with the form[26]

$$
Q = \frac{1}{4\pi}\iint q\,dxdy, \quad q = \vec{m}\cdot\left(\frac{\partial\vec{m}}{\partial x}\times\frac{\partial\vec{m}}{\partial y}\right),
\tag{6}
$$

where $q$ is the topological density. Figures 3(c) and (d) show the topological densities corresponding to the magnetic skyrmions shown in Figs. 3(a) and (b), respectively. It can be seen that $Q = -1$ in Fig. 3(c) and $Q = 1$ in Fig. 3(d), i.e., $Q = -W$ when the spins point down in the central region and point up in the boundary region.

**B. Skyrmion size and dissipative force tensor**

The diameter of twisted skyrmion size ($d$) is usually defined as the distance from in-plane to in-plane magnetization, i.e., the distance between the region $m_z = 0$, as shown in the inset of Fig. 4. The dissipative force tensor $\mathcal{D}$ is used to describe the effect of the dissipative forces on the moving skyrmion[46-48]. For a single twisted skyrmion, $\mathcal{D}$ is given by

$$\mathcal{D} = 4\pi \begin{pmatrix} \mathcal{D} & 0 \\ 0 & \mathcal{D} \end{pmatrix}, \quad \mathcal{D} = \frac{1}{4\pi} \int \frac{\partial \vec{m}}{\partial x} \cdot \frac{\partial \vec{m}}{\partial x} dxdy, \tag{7}$$

where $\mathcal{D}$ is the diagonal element of the dissipative tensor and also called dissipative parameter. The dissipative parameter $\mathcal{D}$ is determined by the diameter and domain wall width of the twisted skyrmion. Therefore, both $d$ and $\mathcal{D}$ are affected by DMI strength as shown in Fig. 4. With the increase in $D_{DMI}$ from 2.5 to 3.5 mJ/m$^2$, $d$ increases from 7.9 to 34.8 nm and $\mathcal{D}$ increases from 1.0577 to 1.961, respectively, for the twisted skyrmion.

**Dynamics of twisted skyrmion driven by the STT**

To understand the STT-induced motion of the twisted skyrmions, we first use the Thiele equation[41] to describe the dynamics of the four kinds of skyrmions mentioned above by casting the LLG Eqs. (1) and (2) to the following equation[46, 47]:

$$\mathbf{G} \times (v_s - v_d) + \mathcal{D}(\beta v_s - \alpha v_d) = 0, \tag{8}$$

where $\mathbf{G}$ is the gyrovector with the form $\mathbf{G} = (0\ 0\ G) = (0\ 0\ 4\pi Q)$, and $v_d$ is the drift velocity of the skyrmion. When the velocity of the conduction electrons $v_s$ applied along the $x$ direction, $v_d = (v_x, v_y)$ is derived from Eq. (8) as

$$\begin{cases} v_x = \left( \dfrac{\beta}{\alpha} + \dfrac{Q^2(\alpha - \beta)}{\alpha(Q^2 + \alpha^2 \mathcal{D}^2)} \right) v_s, \\ v_y = Q \dfrac{\mathcal{D}(\alpha - \beta)}{Q^2 + \alpha^2 \mathcal{D}^2} v_s. \end{cases} \tag{9}$$

It can be seen that the direction of the skyrmion deviates from the direction of the conduction electrons when $\alpha \neq \beta$, and this phenomenon is called the skyrmion Hall effect and can be further defined by the skyrmion Hall angle

$$\theta_{Sky} = \text{sign}(v_y) \cdot \arccos\left( \frac{v_x}{\sqrt{v_x^2 + v_y^2}} \right), \tag{10}$$

which defines the angle in the range from $-180°$ to $180°$. For the situation of STT-induced skyrmion motion, the sign of the $v_x$ is always the same with $v_s$, i.e., the skyrmion Hall angle is in the range of (-90°, 90°), and therefore the Eq. (10) can be reduced to $\theta_{Sky} = \arctan\left( \dfrac{Q\mathcal{D}(\alpha - \beta)}{Q^2 + \alpha\beta\mathcal{D}^2} \right)$.

The trajectories of the four types of skyrmion driven by the in-plane STT with $v_s = 100$ m/s, $\alpha = 0.4$, $\beta = 0.2$ and $D_{\text{DMI}} = 3$ mJ/m$^2$ is shown in Fig. 5. The positions of the skyrmions are obtained by solving the guiding center ($R_x$, $R_y$) with the form[49, 50]

$$R_x = \frac{\iint xq\,dxdy}{\iint q\,dxdy}, \quad R_y = \frac{\iint yq\,dxdy}{\iint q\,dxdy}, \qquad (11)$$

where $q$ is the topological density. One can see that the antiskyrmion deflects to the $+y$ direction, while for Bloch skyrmion, Néel skyrmion and twisted skyrmion deflect to the $-y$ direction, i.e., $\theta_{\text{Sky}}$ of the skyrmions with $Q = 1$ (antiskyrmion) and $Q = -1$ (Bloch, Néel and twisted skyrmion) equal to 12.89° and −12.89°, respectively. Following we focus on the STT-induced motion of twisted skyrmion with different conditions, as shown in Fig. 6. Figures 6 (a) and (b) show the $v_x$ and $v_y$ as a function of $v_s$ for different $\alpha$ with $\beta = 0.2$ and $D_{\text{DMI}} = 3$ mJ/m$^2$, respectively. It can be seen that $v_x$ and $v_y$ both increase linearly with the increase in $v_s$ for different $\alpha$, it should be also note that $v_y$ is a negative value for $\alpha < \beta$, a positive value for $\alpha > \beta$, and zero for $\alpha = \beta$. Then we chose the situation of $v_s = 100$ m/s to investigate the skyrmion Hall angle of the twisted skyrmion as a function of $v_s$, as shown in Fig. 6(c), the skyrmion Hall angel $\theta_{\text{Sky}}$ remains almost unchanged with the increase in $v_s$. Figure 6(d) shows the simulation and calculation of $\theta_{\text{Sky}}$ as a function of $\alpha$ with $\beta = 0.2$, the skyrmion Hall angle $\theta_{\text{Sky}}$ decreases from 13.7° to −12.89° with the $\alpha$ increasing from 0.01 to 0.4. According to Eqs. (9) and (10), both the velocity and the skyrmion Hall angle $\theta_{\text{Sky}}$ are affected by the dissipative parameter $\mathcal{D}$, and the dissipative parameter $\mathcal{D}$ is determined by the DMI strength $D_{\text{DMI}}$. Therefore, it is necessary to investigate the dynamics of the twisted skyrmion under different $D_{\text{DMI}}$, as shown in Figs. 6 (e) and (f) with $v_s = 100$ m/s, $\alpha = 0.4$ and $\beta = 0.2$. $v_x$ increases at first and then decreases with $D_{\text{DMI}}$ increasing from 2.5 to 3.5 mJ/m$^2$, while $v_y$ keeps decreasing (the absolute value of $v_y$ is continuously increasing), and both simulation and calculation results support that the corresponding skyrmion Hall angle $\theta_{\text{Sky}}$ decreases from −11.4° to −16.8° (the absolute value of $\theta_{\text{Sky}}$ is proportional to the $D_{\text{DMI}}$).

We have known that the STT-induced twisted skyrmion motion is affected by the damping in the previous paragraph. Following, we investigate the dynamics of twisted skyrmion induced by the STT under a damping gradient, as shown in Fig. 7. Figure 7(a) shows the position along the $y$ axis of the twisted skyrmion as a function of distance along the $x$ axis with $v_s$ = 100 m/s, $\beta$ = 0.4 and $D_{DMI}$ = 3 mJ/m$^2$. The damping decreases from 0.5 to 0.25 linearly from 0 to 50 nm along the *x* axis, as indicated by the color code. Figure 7 (b) shows the skyrmion Hall angle $\theta_{Sky}$ of the twisted skyrmion as a function of its position along the *x* axis. In the region $\alpha > \beta$, the twisted skyrmion moves along the *x* axis direction from 0 nm and deflects in the –*y* direction until moving to the *x* axis of 20 nm, where $\alpha = \beta = 0.4$; from the region of 20 to 50 nm along *x* axis, the twisted skyrmion begins to deflect in the +*y* direction because of $\alpha < \beta$. Therefore, the trajectory of twisted skyrmion induced by the STT can be controlled under a damping gradient.

**Dynamics of twisted skyrmion driven by the SHE**

SHE-induced motion of antiskyrmion has already been studied in Ref. [38], which demonstrates that the antiskyrmion Hall angle depends on the direction of the current strongly. In this section, we focus on the SHE-induced motions of the skyrmions whose winding number *W* = 1(Bloch, Néel and twisted skyrmion). The LLG Eqs. (1) and (3) can be cast into the following form:

$$\mathbf{G} \times v_d + \alpha \mathcal{D} v_d - 4\pi B \mathbf{R}(\chi) J_{HM} = 0 \tag{12}$$

where $\mathbf{G}$ = (0 0 –4$\pi$) due to *Q* = –1, *B* is linked to the SHE, and the sign of *B* is determined by the SHE angle; $\mathbf{R}(\chi)$ is the in-plane rotation matrix with the form $\mathbf{R}(\chi) = \begin{pmatrix} \cos\chi & \sin\chi \\ -\sin\chi & \cos\chi \end{pmatrix}$ [49, 51]. When the current $J_{HM}$ injected into the heavy metal along the *x* direction, $v_d = (v_x, v_y)$ is derived from Eq. (12) as

$$\begin{cases} v_x = B \dfrac{\alpha \mathcal{D} \cos\chi + \sin\chi}{1+\alpha^2 \mathcal{D}^2} J_{HM}, \\ v_y = B \dfrac{-\alpha \mathcal{D} \sin\chi + \cos\chi}{1+\alpha^2 \mathcal{D}^2} J_{HM}. \end{cases} \tag{13}$$

The skyrmion Hall angle $\theta_{Sky}$ can be obtained by the Eq. (10), which is in the range of –180° to 180°.

The Eq. (13) suggests that the direction of motion of the skyrmions depends on their helicities. Therefore, we first investigate the trajectories of skyrmions driven by the SHE with $J_{HM}$ = 10 × 10$^{10}$ A/m$^2$, $\alpha$ = 0.2 and $D_{DMI}$ = 3 mJ/m$^2$ for different helicities of skyrmions, as shown in Fig. 8. These skyrmions with different helicities are achieved by changing the direction of DMI vector. The simulation results in Fig. 8(a) show that the skyrmion Hall angles $\theta_{Sky}$ are –150.4°,

165.4°, 121.4°, 75.6°, 29.6°, −14.6°, −58.6° and −104.4° for the helicities $\chi = -0.75\pi, -0.5\pi, -0.25\pi, 0, 0.25\pi, 0.5\pi, 0.75\pi$ and $\pi$, respectively. Figure 8(b) shows the skyrmion Hall angle as a function of the helicity both supported by simulations and calculations. Following we take the case of $\chi = 0.25\pi$ (the twisted skyrmion shown in Fig. 1(d)) and investigate the motion induced by the SHE, as shown in Fig. 9. Figure 9(a) shows the simulation results of $v_x$ and $v_y$ of the twisted skyrmion as a function of $J_{HM}$ with $\alpha = 0.2$ and $D_{DMI} = 3$ mJ/m$^2$. It can be seen that $v_x$ and $v_y$ both increase linearly with the increase in $J_{HM}$, and the corresponding skyrmion Hall angle $\theta_{Sky}$ is shown in Fig. 9(b). The skyrmion Hall angle $\theta_{Sky}$ almost remains at 29.6° when $J_{HM}$ is no more than $200 \times 10^{10}$ A/m$^2$, while for the case $J_{HM} = 500 \times 10^{10}$ A/m$^2$, the skyrmion Hall angle $\theta_{Sky}$ decreases to 28.9°. This is because the size of the twisted skyrmion, i.e., the dissipative parameter $\mathcal{D}$, increases slightly with $J_{HM}$ increasing to $500 \times 10^{10}$ A/m$^2$, the skyrmion Hall angle Eq. (10) can be reduced to

$$\theta_{Sky} = \arctan(\frac{1-\alpha\mathcal{D}}{1+\alpha\mathcal{D}}). \qquad (14)$$

For $\chi = 0.25\pi$, which indicates that the skyrmion Hall angle $\theta_{Sky}$ decreases with the increase in $\mathcal{D}$. Figure 9(c) shows the simulation results of $v_x$ and $v_y$ of the twisted skyrmion as a function of $\alpha$ with $J_{HM} = 100 \times 10^{10}$ A/m$^2$ and $D_{DMI} = 3$ mJ/m$^2$, $v_x$ first increases and then decreases with $\alpha$ increasing from 0.01 to 1, while $v_y$ keeps decreasing (the absolute value of $v_y$ decreases at first and then increases), and therefore the corresponding skyrmion Hall angle $\theta_{Sky}$ decreases from 44.3° to −8.2° (the trend of $\theta_{Sky}$ is consistent with $v_y$), which also supported by calculation, as shown in Fig. 9(d). Figure 9(e) shows that $v_x$ and $v_y$ both increases with $D_{DMI}$ increasing from 2.5 to 3.5 mJ/m$^2$ when $J_{HM} = 100 \times 10^{10}$ A/m$^2$ and $\alpha = 0.2$. Figure 9(f) shows that the corresponding skyrmion Hall angle $\theta_{Sky}$ decreases with the increase in $D_{DMI}$, which is similar to the results by calculating the Eq. (14) with the increase in $\mathcal{D}$.

In contrast to the STT-induced twisted skyrmion motion under a damping gradient, we investigate the dynamics of twisted skyrmion driven by the SHE under a damping gradient, as shown in Fig. 10. Figure 10(a) shows the trajectory of the twisted skyrmion as a function of its position along the $x$ axis with $J_{HM} = 100 \times 10^{10}$ A/m$^2$ and $D_{DMI} = 3$ mJ/m$^2$. The damping increases from 0.2 to 1.2 linearly from 0 to 200 nm along the $x$ axis, as indicated by the color code. Figure 10 (b) shows the corresponding skyrmion Hall angle $\theta_{Sky}$ as a function of its position along the *x* axis. The Eq. (14) implies that: in the region $1-\alpha\mathcal{D} > 0$, the twisted skyrmion moves along the *x* axis direction from 0 nm and deflects in the +*y* direction until moving to the *x* axis of 114 nm where $1-\alpha\mathcal{D} = 0$; in the region $1-\alpha\mathcal{D} < 0$, i.e., from 114 to 200 nm along *x* axis, the twisted skyrmion deflects in the −*y* direction. Therefore, the trajectory of the SHE-induced motion of twisted skyrmion can also be controlled by a damping gradient.

**Conclusions**

In summary, we first introduce the magnetic structure and the corresponding DMI of the twisted skyrmion in contrast to that of Bloch skyrmion, Néel skyrmion and antiskyrmion. Furthermore, we discuss and calculate the helicity, winding number, topological number, size and dissipative force tensor of the twisted skyrmion, which pave the way for the following study of the dynamics of twisted skyrmion driven by the STT and the SHE. For the STT-induced motion of twisted skyrmion, it is found that the skyrmion Hall angle is determined by the topological number, the dissipative force tensor and the difference between the Gilbert damping and the non-adiabatic factor. For the SHE-induced motion of twisted skyrmion, apart from the dissipative force tensor and the Gilbert damping, the skyrmion angle depends on the helicity significantly. At last, we demonstrate that the trajectories of both the STT-induced and the SHE-induced motion of twisted skyrmion can be controlled by a Gilbert damping gradient. These results may present guidance for the design of twisted skyrmion-based racetrack memories.

**Acknowledgments**

This work is supported by National Science Fund of China (11574121 and 51771086). C. J. acknowledges the funding by the China Scholarship Council.

**Figures**

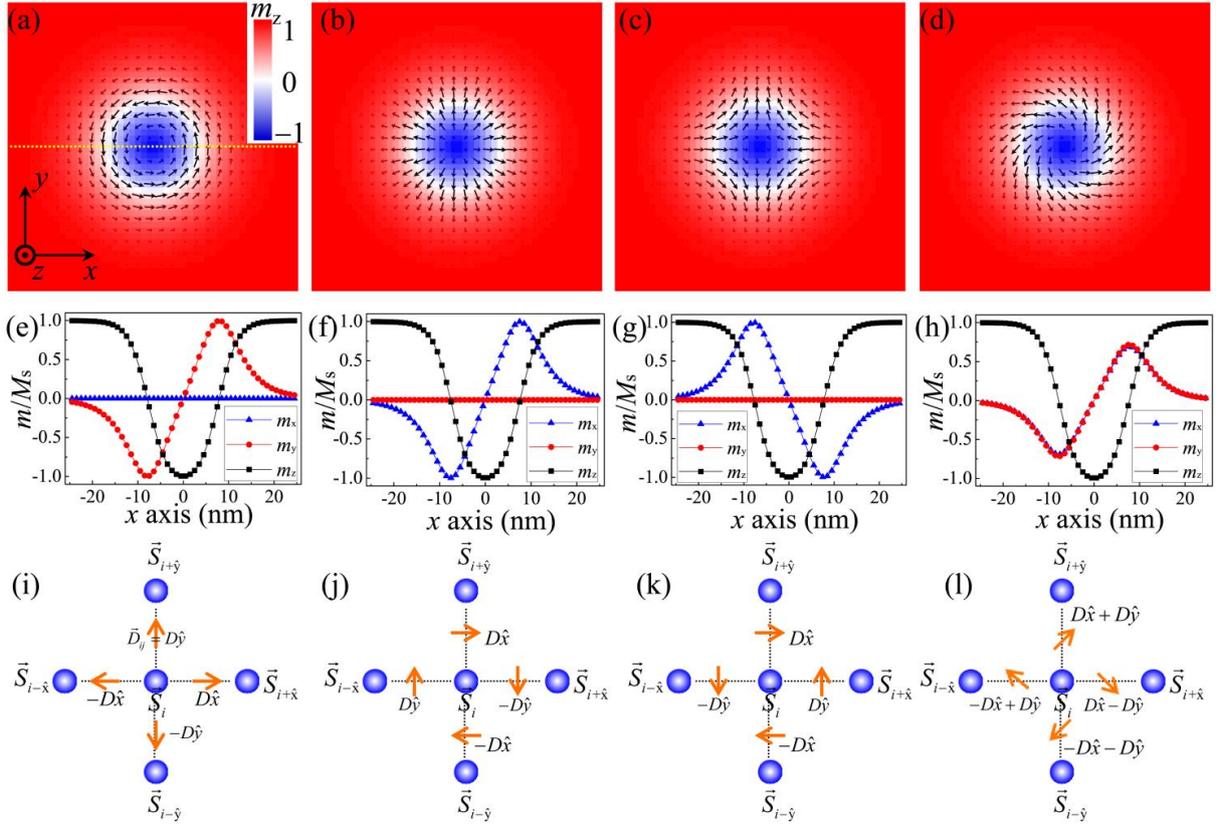

FIG. 1. Four types of skyrmions. (a)–(d) display the magnetization distribution of Bloch skyrmion, Néel skyrmion, antiskyrmion and twisted skyrmion, respectively. We only intercept the central region of the 2D plane with the size of 50 nm × 50 nm. The red, white and blue represent where the $z$ component of the magnetization is positive, zero and negative, respectively. The black arrows denote the distribution of the in-plane magnetization. (e)–(h) are the spatial profiles of the local magnetization corresponding to the yellow dotted line which marked in the Fig. 1(a). (i)–(l) are the configurations of bulk DMI, interfacial DMI, anisotropic DMI and twisted DMI, respectively. The orange arrows denote the directions of the DMI vector.

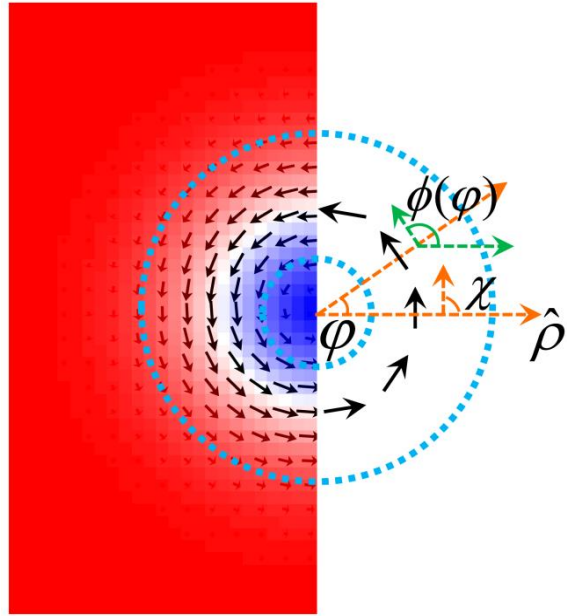

FIG. 2. Schematic of a general skyrmion in two-dimensional polar coordinates. $\rho$, $\varphi$, $\chi$ and $\phi(\varphi)$ indicate the radial coordinate, azimuthal angle, skyrmion helicity and the angle between the magnetic moment and the radial coordinate, respectively.

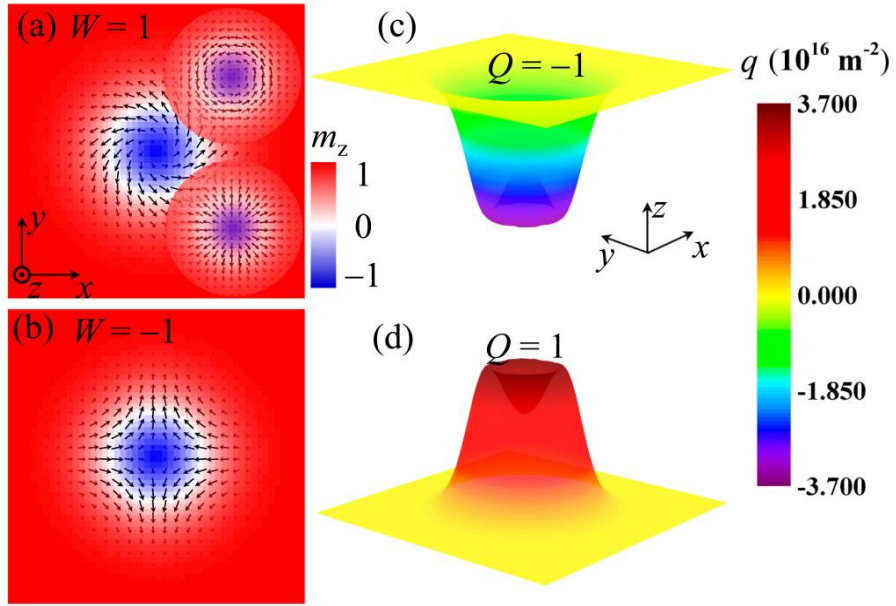

FIG. 3. (a) displays the magnetization distributions of twisted skyrmion, Bloch skyrmion and Néel skyrmion with $W = 1$. (b) displays the magnetization distribution of antiskyrmion with $W = -1$. (c) and (d) show the distributions of topological density corresponding to the magnetizations shown in (a) and (b) with $Q = -1$ and $Q = 1$, respectively.

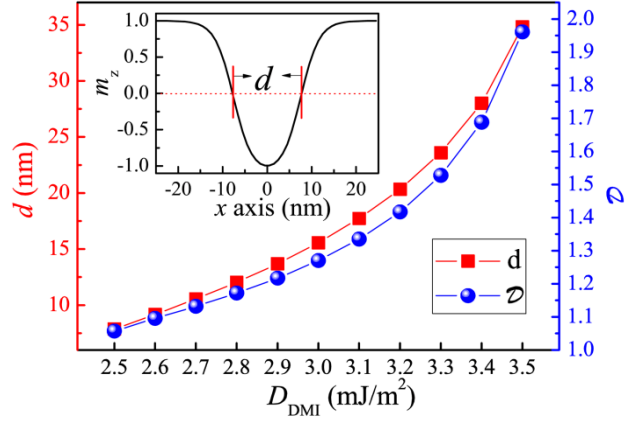

FIG. 4. Skyrmion diameter ($d$) and the diagonal element of the dissipative tensor ($\mathcal{D}$) as a function of DMI strength. The inset is the spatial profile of $m_z$ across the twisted skyrmion. It should be note that the twisted skyrmion exists stably in region of 250 nm ×250 nm, the diagram only show the central part of 50 nm ×50 nm.

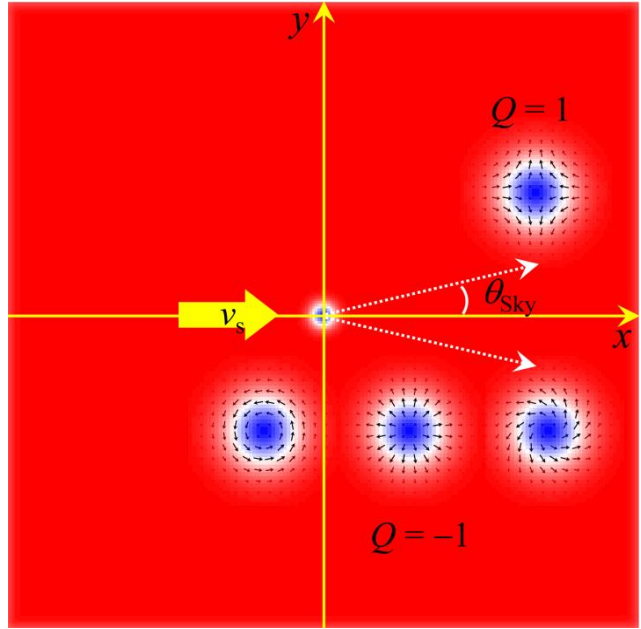

FIG. 5. The trajectories of four types of skyrmion driven by the STT. The initial position of the skyrmions is at the center of the 2D magnetic film, the size of 2D plane is 250 nm × 250 nm, $v_s$ = 100 m/s in $x$ direction, $\alpha$ = 0.4, $\beta$ = 0.2 and $D_{DMI}$ = 3 mJ/m$^2$. The big yellow solid arrow and white dotted arrows represent the direction of conduction electrons and the trajectories of skyrmions, respectively. It should be note that the four types of skyrmions are enlarged with the purpose to see their helicities clearly. The actual sizes of the four skyrmions are almost the same as the skyrmion at the center position.

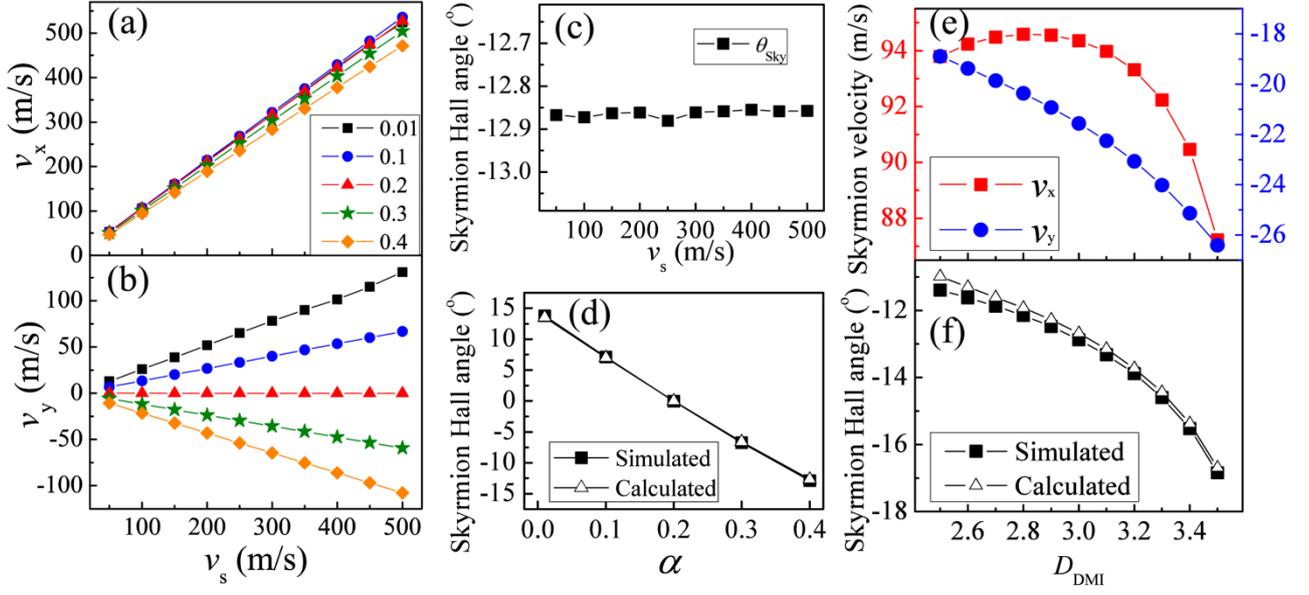

FIG. 6. The STT-induced motion of the twisted skyrmion ($\chi = 0.25\pi$). (a) and (b) display the $v_x$ and $v_y$ as a function of $v_s$ for $\alpha = 0.01$, 0.1, 0.2, 0.3 and 0.4 with $\beta = 0.2$ and $D_{DMI} = 3$ mJ/m$^2$, respectively. (c) The skyrmion Hall angle $\theta_{Sky}$ as a function of $v_s$ corresponding to the situation of $\alpha = 0.4$ shown in Figs. (a) and (b). (d) The skyrmion Hall angle $\theta_{Sky}$ as a function of $\alpha$ corresponding to the situation of $v_s = 100$ m/s shown in Figs. 6 (a) and (b). (e) and (f) display the skyrmon velocity and the skyrmion Hall angle as a function of $D_{DMI}$ with $v_s = 100$ m/s, $\alpha = 0.4$ and $\beta = 0.2$, respectively.

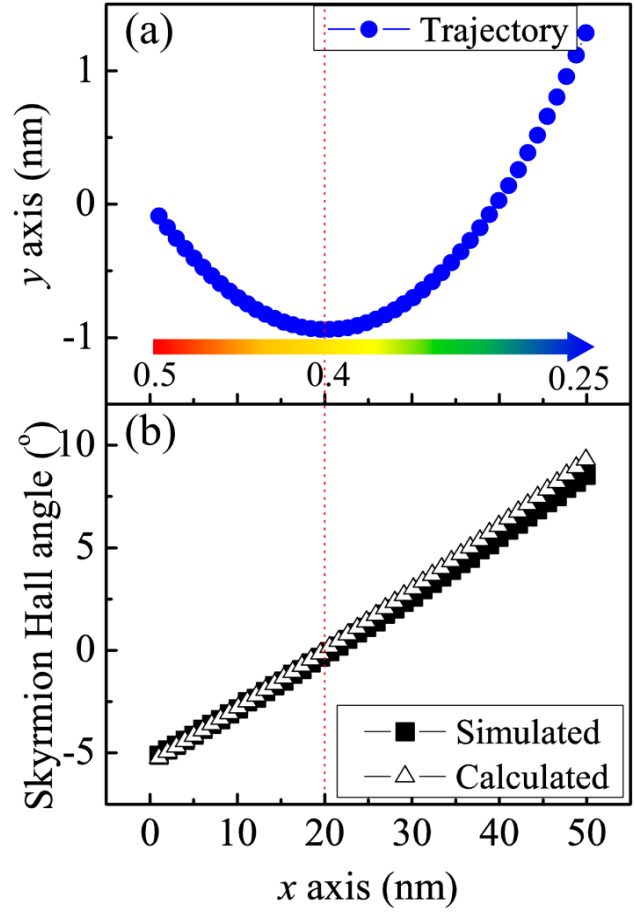

FIG. 7. The STT-induced motion of the twisted skyrmion under a damping gradient. (a) and (b) show the transverse distance (y axis) of the skyrmion and the corresponding skyrmion Hall angle $\theta_{Sky}$ as a function of radial distance (x axis), respectively. The initial position of the skyrmions is defined as 0 nm both in x and y axis, $v_s$ = 100 m/s, $\beta$ = 0.4 and $D_{DMI}$ = 3 mJ/m$^2$. The color code represents that the damping $\alpha$ decreases from 0.5 to 0.25 linearly in the region from 0 to 50 nm along the x direction. The red dotted line represents the position where $\alpha = \beta$.

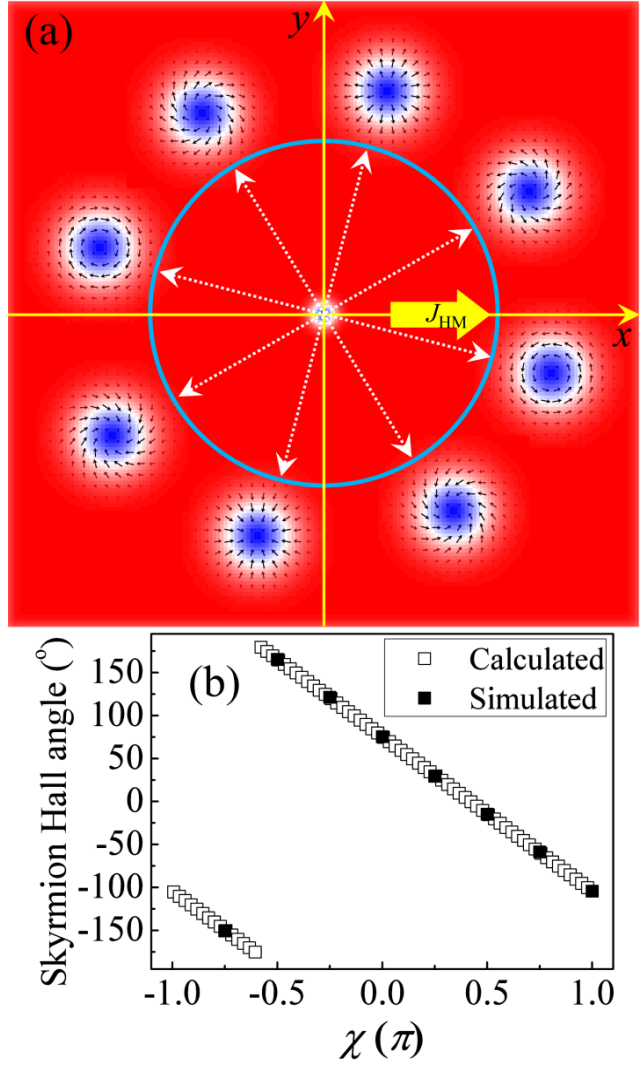

FIG. 8. The SHE-induced motion of skyrmions with different $\chi$ (a) The trajectories of eight types of skyrmions with $\chi = -0.75\pi, -0.5\pi, -0.25\pi, 0, 0.25\pi, 0.5\pi, 0.75\pi$ and $\pi$ driven by the SHE. The initial position of the eight skyrmions is in the center of the 2D magnetic film whose size is 250 nm × 250 nm, $\alpha = 0.2$ and $D_{DMI} = 3$ mJ/m$^2$. The big yellow solid arrow denotes the direction of current $J_{HM} = 10 \times 10^{10}$ A/m$^2$. The white dotted arrows represent the trajectories of skyrmions. It also should be note here that the eight types of skyrmions are enlarged to see their helicities clearly. The actual sizes of the eight skyrmions are almost the same as them at the center position. (b) The skyrmion Hall angle $\theta_{Sky}$ as a function of the helicitiy $\chi$. The black solid squares correspond to the eight types of skyrmion in Fig. 8(a), and the black hollow squares are calculated by the equation.

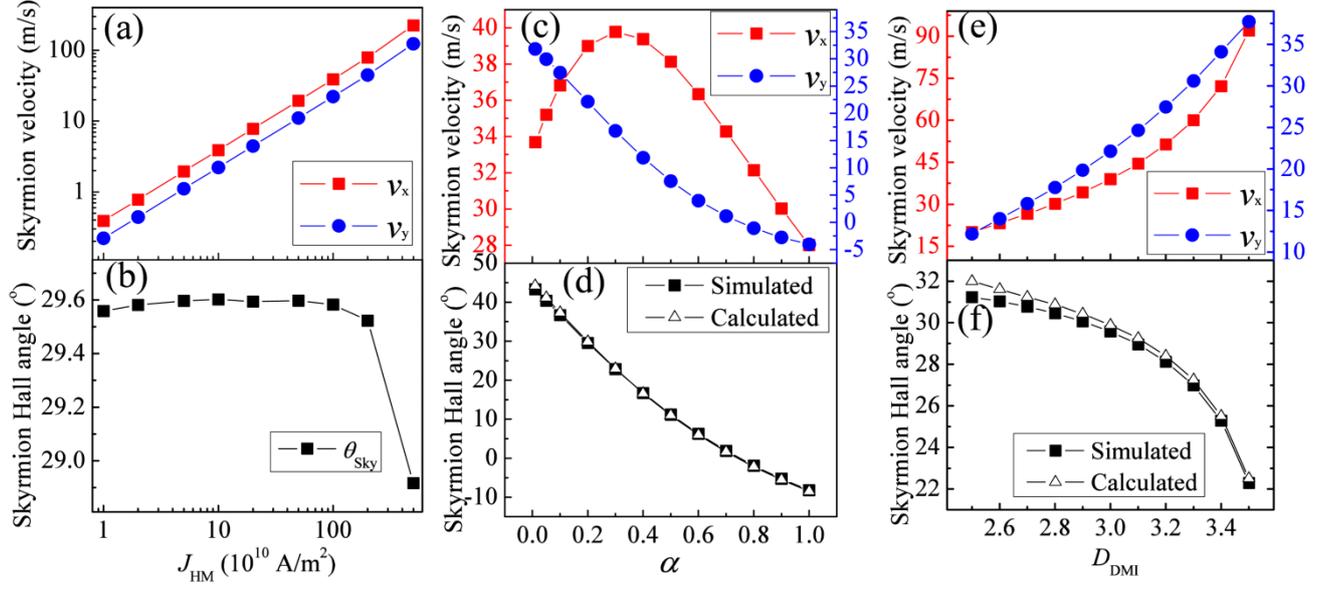

FIG. 9. The SHE-induced motion of the twisted skyrmion ($\chi = 0.25\pi$). (a) and (b) display the skyrmion velocity and skyrmion Hall angle $\theta_{Sky}$ as a function of $J_{HM}$ with $\alpha = 0.2$ and $D_{DMI} = 3$ mJ/m$^2$, respectively. (c) and (d) denote the skyrmion velocity and skyrmion Hall angle $\theta_{Sky}$ as a function of $\alpha$ with $J_{HM} = 100 \times 10^{10}$ A/m$^2$ and $D_{DMI} = 3$ mJ/m$^2$, respectively. (e) and (f) represent the skyrmion velocity and skyrmion Hall angle $\theta_{Sky}$ as a function of $D_{DMI}$ with $J_{HM} = 100 \times 10^{10}$ A/m$^2$ and $\alpha = 0.2$, respectively.

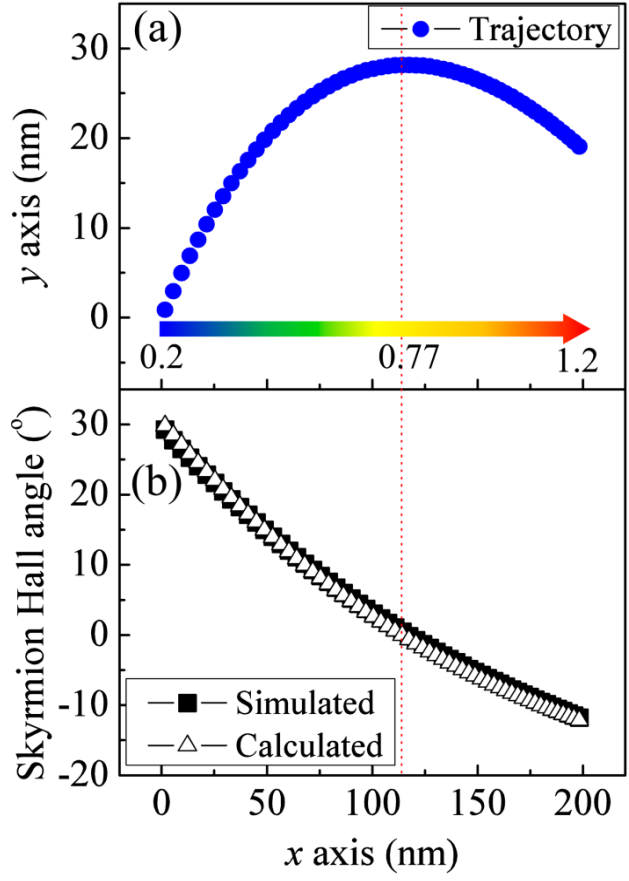

FIG. 10. The SHE-induced motion of the twisted skyrmion under a damping gradient. (a) and (b) show the transverse distance (*y* axis) of the skyrmion and the corresponding skyrmion Hall angle $\theta_{Sky}$ as a function of radial distance (*x* axis), respectively. The initial position of the skyrmions is defined as 0 nm both in *x* and *y* axis, $J_{HM} = 100 \times 10^{10}$ A/m$^2$ and $D_{DMI} = 3$ mJ/m$^2$. The color code represents that the damping $\alpha$ increases from 0.2 to 1.2 linearly in the region from 0 to 200 nm along the *x* direction. The red dotted line represents the position where $1-\alpha\mathcal{D} = 0$, i.e., the skyrmion Hall angle $\theta_{Sky} = 0°$.